\documentclass
[twocolumn,showpacs,preprintnumbers,amsmath,amssymb]{revtex4}

\usepackage{graphicx}
\usepackage{dcolumn}
\usepackage{bm}
\usepackage{rotating}
\usepackage{color}
\usepackage{soul}
\usepackage{bigints}

\begin{document}
\title{Wheeler equation in Kerr background\\and astrophysical jet collimation}
\date{\today}
\author{X. H. Zheng$^1$}
\email{xhz@qub.ac.uk} 
\author{J. X. Zheng$^2$} 
\affiliation{$^1$Department of Physics, Queen's University of Belfast, BT7 1NN, Northern Ireland}
\affiliation{$^2$Department of Electrical and Electronic Engineering, Imperial College London, SW7 2AZ, England}

\begin{abstract}
The Wheeler equation has an extension in the Kerr background.  It has an exact solution representing a running harmonic wave in the tortoise coordinate.  The wave models astrophysical jets in a new light because it collimates along the axis of spacetime rotation, increasingly tightly when rotation quickens.  The model endorses a previous numerical work suggesting intrinsic collimation of jet fuelled by a Penrose-like process.  It also endorses a previous theoretical work to crop one-way running waves from the exact solutions to Heun's equation.
\end{abstract}
\pacs{general relativity, Wheeler equation, astrophysical jet collimation}
\maketitle

\section{Introduction}\label{sec:1}
There have been two extensive lines of investigations to understand astrophysical jet collimation.  Both are becoming increasingly quantitative with observational and experimental evidence.  In 1978 Rees discussed the supply of gas to sustain the process of accretion to fuel electromagnetic radiation \cite{Rees}.  In 1982 Blandford and Payne examined the possibility that energy and angular momentum are removed magnetically from accretion discs by field lines that leave the disc surface and extend far \cite{Blandford}.  In 2021 Revet and co-workers reported laboratory disruption of scaled astrophysical outflows by a misaligned magnetic field.  In their experiment, an expanding plasma outflow, generated by ablating plasma from a solid by a laser, interacts with a static magnetic field.  They observed a collimated plasma jet when they applied the magnetic field \cite{Revet}.

In the other line of investigation, assuming a Penrose-like process to fuel the jets, collimation arises from spacetime rotation.  In 2004, Williams found from theoretical analysis and Monte Carlo simulation the helical polar angles of the jet particles range from $40^\circ$ to $0.5^\circ$ (for the highest energy particles) \cite{Williams}.  In 2010 Gariel and co-workers re-examined the topic and found a relation between the asymptotes of the outflow jet and mass, energy, charge and rate of rotation of the black hole \cite{Gariel}.

Theoretically, back to 1972, Brill and co-workers found an equation for electromagnetic disturbances in the Kerr background \cite{Brill}.  Shortly afterwards, also in 1972, Teukolsky derived a slightly more general version of that equation \cite{Teukolsky}.  In 1998, Suzuki, Takasugi and Umetsu transformed the Teukolsky equation into Heun's equation \cite{Suzuki}.  In 2010, Fiziev presented a unified description of the exact solutions to the confluent version of Heun's equation.  He found a proper linear combination of such solutions represents bounded one-way running waves \cite{Fiziev2}.

Fiziev expected the bounded one-way running waves he cropped up for the first time could describe the real astrophysical jets, observed at very different scales in the Universe.  He acknowledged the idea is still speculative and called for a more detailed mathematical development and a careful confrontation with the actual astrophysical observations \cite{Fiziev2}.

To answer the call, we trace the line back to the good old days when the pursuit for completeness and perfection had not yet clouded physics.  In 1955 John Archibald Wheeler derived an elegant equation for intensive electromagnetic radiation capable of distorting spacetime to trap itself to form entities called geons \cite{Wheeler}.  In 1957, Regge and Wheeler rederived the Wheeler equation to include gravitational perturbations \cite{Regge}.  In 2006, Fiziev specified the Regge-Wheeler equation with the Schwarzschild metric and transformed it into Heun's equation for exact solutions \cite{Fiziev}.

But the Regge-Wheeler equation already has an exact solution.  It assumes a specific polarization for the electromagnetic radiation it describes and thus becomes simpler.  It resembles the Helmholtz equation because Wheeler introduced a factor to cancel the geometric effect of wavefront expansion.  Its solution can be exact, reminiscent of a harmonic wave but in the tortoise coordinate in the Schwarzschild background.

We extend the Wheeler equation to the Kerr background.  The radiation assumes the original Wheeler polarisation for simplicity.  We pay special attention to the non-diagnostic components of the metric, which are absent in the original Wheeler formalism.  The extended Wheeler equation also has an exact solution, reminiscent of a harmonic wave but in the tortoise coordinate in the Kerr background.  It represents a beam of radiation collimated along the axis of rotation, tighter when the black hole rotates quicker (or when the frequency of radiation increases).

We find the velocity of light in the tortoise coordinates is a proper velocity, displacement measured by a remote observer but time elapsed on the local clock.  We find space and time always play the usual roles across the metric.  The radiation must be running waves supported by a source.  It cannot be standing waves to serve as a functional basis to present any electromagnetic disturbances with a Fourier-styled expansion.

We arrange our derivation, narration and discussion as follows.  In Section~\ref{sec:2} we outline the steps to derive the Wheeler equation in the Schwarzschild background.  In Section~\ref{sec:3} we extend the Wheeler equation to the Kerr background.  In Section~\ref{sec:4} we show the relationship between the Wheeler equation and the equations of Bessel, Legendre, Brill and others.  In Section~\ref{sec:5} we show that, in the tortoise coordinate, the speed of light is a proper velocity.  In Section~\ref{sec:6} we show space and time play the usual roles across both the Schwarzschild and Kerr metrics.  In Section~\ref{sec:7} we show radiation in the Schwarzschild and Kerr background must be running waves.  We place brief conclusions in Sections~\ref{sec:8}.

\section{Wheeler equation in Schwarzschild background}\label{sec:2}
\begin{figure}[t]
\centering\includegraphics[width=9cm]{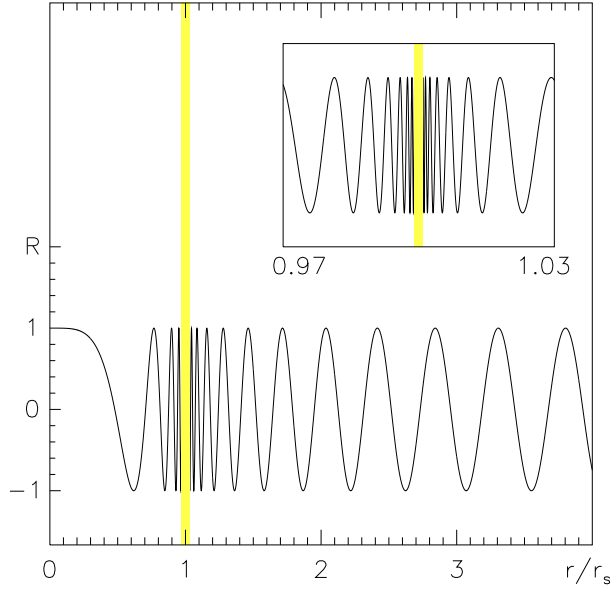}
\caption{Real part of the solution to the Wheeler equation in Schwarzschild background, $\Omega r_s = 9.1$ and $C = 0$.  For clarity, a section (width = 0.06 in $r_s$) is shaded and shown in the inset, where a section also is shaded. There can be an infinite number of such zoom-in presentations, like a Russian doll with infinite layers.}
\label{fig:1}
\end{figure}

The Schwarzschild metric \cite{Schwarzschild} is of the form 
\begin{eqnarray}\label{eq:1}
ds^2 = g_{\alpha\beta}\,dx^\alpha dx^\beta
\end{eqnarray}
which leads to the contravariant metric:
\begin{eqnarray}
\frac{d^2}{ds^2} = g^{\alpha\beta}\frac{\partial}{\partial x^\alpha}\frac{\partial}{\partial x^\beta}
\end{eqnarray}
with
\begin{eqnarray}\label{eq:3}
\begin{array}{lll}
g_{rr} = e^\lambda,&&g^{rr} = e^{-\lambda},\\
g_{\theta\theta} = r^2,&&g^{\theta\theta} = r^{-2},\\
g_{\varphi\varphi} = r^2\sin^2\theta,&&g^{\varphi\varphi} = 1/r^2\sin^2\theta,\\
g_{TT} = -e^\nu,&&g^{TT} = -e^{-\nu},
\end{array}
\end{eqnarray}
$T = ct$, and
\begin{eqnarray}\label{eq:4}
e^\lambda = \left(1 - \frac{r_s}{r}\right)^{-1}\;\;\;\;\mbox{and}\;\;\;\;e^\nu = 1 - \frac{r_s}{r}
\end{eqnarray}
where $r_s$ is the Schwarzschild radius in association with the mass of the black hole.  In Wheeler's treatment \cite{Wheeler} 
\begin{eqnarray}\label{eq:5}
\frac{1}{\sqrt{-g}}\,\frac{\partial}{\partial x^\alpha}\left(\sqrt{-g}\,\frac{g^{\alpha\beta}}{r^2\sin^2\theta}\,\frac{\partial A}{\partial x^\beta}\right) = 0
\end{eqnarray}
assuming the electromagnetic disturbance is source-free, $g = -r^4\sin^2\theta$ being the determinant of $g_{\alpha\beta}$,
\begin{eqnarray}\label{eq:6}
&&\frac{e^{-\lambda}}{r^2\sin^2\theta}\,\frac{\partial A}{\partial r}\!\sim\!-H_\theta,\nonumber\\
&&\frac{1}{r^2\sin^2\theta}\,\frac{\partial A}{\partial\theta}\sim\;H_r,\\
&&\frac{-e^{-\nu}}{r^2\sin^2\theta}\,\frac{\partial A}{\partial T}\sim\;E_\varphi\nonumber
\end{eqnarray}
components of the electromagnetic field, with just one electrical component in the $\varphi$-direction.  In Eqs.~(\ref{eq:5}) and (\ref{eq:6}) Wheeler introduced a factor, $1/r^2\sin^2\theta$, to modify the dependence of $H_\theta$, $H_r$ and $E_\varphi$ on $r$ and $\theta$.  He found
\begin{eqnarray}\label{eq:7}
r^2\frac{\partial}{\partial r}\left(e^\lambda\frac{\partial A}{\partial r}\right) + \sin\theta\frac{\partial}{\partial\theta}\left(\frac{1}{\sin\theta}\frac{\partial A}{\partial\theta}\right)\;\;\;\;\nonumber\\
+\frac{1}{\sin^2\theta}\,\frac{\partial^2A}{\partial\varphi^2} - r^2 e^{-\nu}\frac{\partial^2A}{\partial T^2} = 0
\end{eqnarray}
which is separable.  Letting
\begin{eqnarray}\label{eq:8}
A = R(r)\,\Theta(\theta)\,e^{im\varphi}e^{i\Omega T},
\end{eqnarray}
$\Omega = \omega/c$, we have
\begin{eqnarray}\label{eq:9}
e^\nu\frac{d}{dr}\left(\!e^{-\lambda}\frac{dR}{dr}\right) + \left(\Omega^2 - \frac{C}{r^2}e^\nu\right)R = 0,\end{eqnarray}
\begin{eqnarray}\label{eq:10}
\sin\theta\frac{d}{d\theta}\left(\!\frac{1}{\sin\theta}\frac{d\Theta}{d\theta}\right) + \left(C - \frac{m^2}{\sin^2\theta}\right)\Theta = 0
\end{eqnarray}
to determine the radiation field in the radial and angular directions, respectively, $C$ being the separation constant.  We can write Eqs.~(\ref{eq:9}) and (\ref{eq:10}) as
\begin{eqnarray}\label{eq:11}
\frac{d^2R}{dx^2} + \left(\Omega^2 - \frac{C}{r^2}e^\nu\right)\!R = 0,
\end{eqnarray}
\begin{eqnarray}\label{eq:12}
(1 - z^2)\frac{d^2\Theta}{dz^2} + \left(C - \frac{m^2}{1 - z^2}\right)\Theta = 0
\end{eqnarray}
where $z = \cos\theta$,
\begin{eqnarray}\label{eq:13}
x = r + r_s\ln\bigg|1 - \frac{r}{r_s}\bigg|
\end{eqnarray}
is the tortoise coordinate \cite{Fiziev} formulated to let $x = 0$ when $r = 0$.  Eq.~(\ref{eq:11}) reduces to the Riccati-Bessel equation \cite{Abramowitz} in flat spacetime ($r_s\rightarrow 0$, $x\rightarrow r$, $e^\nu\rightarrow1$).

If we let $C = 0$ then Eq.~(\ref{eq:11}) further reduces to the Helmholtz equation with a solution
\begin{eqnarray}\label{eq:14}
R(r) = \exp(-i\Omega x)
\end{eqnarray}
which is simple and exact.  If we further let $m = 0$ in Eq.~(\ref{eq:8}) then Eq.~(\ref{eq:12}) reduces to $d^2\Theta/dz^2 = 0$ to give
\begin{eqnarray}\label{eq:15}
\Theta(\theta) = 1
\end{eqnarray}
to exclude any phase change, save the change in $r$ that is the radiation is in the radial direction. Eqs.~(\ref{eq:14}) and (\ref{eq:15}) represent a spherical wave running away from $x = 0$. In FIG.~1 we plot the real part of Eq.~(\ref{eq:14}). 

\section{Wheeler equation in Kerr background}\label{sec:3}
\begin{figure}[t]
\centering\includegraphics[width=9cm]{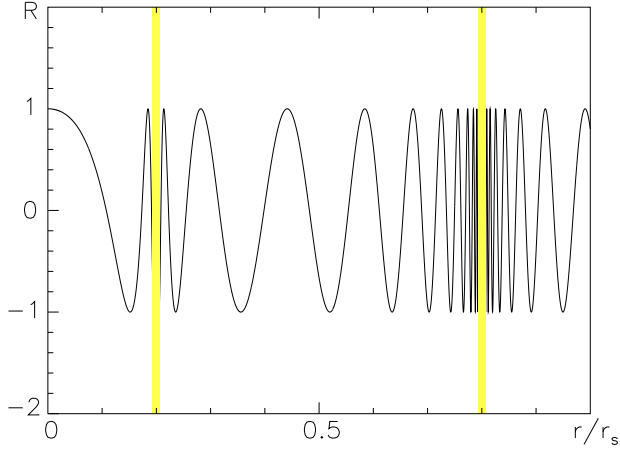}
\caption{Real part of the solution to the Wheeler equation in the Kerr background, $\Omega r_s = 9.1$, $r_- = 0.2$, $r_+ = 0.8$ in $r_s$ from $C = 0$, $m = 0$ and $a = 0.4$.  Sections around $r = r_-$ and $r_+$ are shaded to cover rapid oscillations for clarity.}
\label{fig:2}
\end{figure}

In the Kerr background \cite{Kerr} we replace Eq.~(\ref{eq:3}) with:
\begin{eqnarray}\label{eq:16}
\begin{array}{l}
g_{rr} = \rho^2/\Delta,\\
g_{\theta\theta} = \rho^2,\\
g_{\varphi\varphi} = (r^2+a^2)\sin^2\theta + r_sra^2\sin^4\theta,\\
g_{\varphi T} = -(r_sra/\rho^2)\sin^2\theta,\\
g_{TT} = -(1 - r_sr/\rho^2),
\end{array}
\end{eqnarray}
and
\begin{eqnarray}\label{eq:17}
\begin{array}{l}
g^{rr} = \Delta/\rho^2,\\
g^{\theta\theta} = \rho^{-2},\\
g^{\varphi\varphi} = 1/\rho^2\sin^2\theta - a^2/\rho^2\Delta,\\
g^{\varphi T} = -r_sra/\rho^2\Delta,\\
g^{TT} = -(r^2 + a^2)^2/\rho^2\Delta + (a^2/\rho^2)\sin^2\theta
\end{array}
\end{eqnarray}
where $a$ measures angular momentum and is in length scale,
\begin{eqnarray}\label{eq:18}
\rho^2 = r^2 + a^2\cos\theta\;\;\;\;\mbox{and}\;\;\;\;\Delta = r^2 - r_sr + a^2,
\end{eqnarray}
giving $g = -\rho^4\sin^2\theta$.  We replace Eq.~(\ref{eq:5}) with
\begin{eqnarray}\label{eq:19}
\frac{1}{\sqrt{-g}}\,\frac{\partial}{\partial x^\alpha}\left[\sqrt{-g}\,\frac{g^{\alpha\beta}}{(r^2 + a^2)\sin^2\theta}\,\frac{\partial A}{\partial x^\beta}\right] = 0,
\end{eqnarray}
which leads through Eq.~(\ref{eq:17}) to
\begin{eqnarray}\label{eq:20}
&&(r^2 + a^2)\frac{\partial}{\partial r}\left(\frac{\Delta}{r^2 + a^2}\frac{\partial A}{\partial r}\right) + \sin\theta\frac{\partial}{\partial\theta}\left(\frac{1}{\sin\theta}\frac{\partial A}{\partial\theta}\right)\nonumber\\
&&+\left(\frac{1}{\sin^2\theta} - \frac{a^2}{\Delta}\right)\frac{\partial^2A}{\partial\varphi^2} - \frac{2r_sra}{\Delta}\frac{\partial^2A}{\partial\varphi\partial T}\nonumber\\
&&-\left[\frac{(r^2 + a^2)^2}{\Delta} - a^2\sin^2\theta\right]\frac{\partial^2A}{\partial T^2} = 0
\end{eqnarray}
to replace Eq.~(\ref{eq:7}).  By substituting Eq.~(\ref{eq:8}) into Eq.~(\ref{eq:20}) and separating the variables, we find
\begin{eqnarray}\label{eq:21}
\frac{d^2R}{dy^2} &+& \left(\Omega^2 - \frac{C}{r^2 + a^2}\frac{\Delta}{r^2 + a^2}\right.\nonumber\\
&+&\left.\frac{ma}{r^2 + a^2}\frac{ma - 2r_sr\Omega}{r^2 + a^2}\right)R = 0,
\end{eqnarray}
and
\begin{eqnarray}\label{eq:22}
(1 - z^2)\frac{d^2\Theta}{dz^2} &+& \bigg[C - \frac{m^2}{1 - z^2}\nonumber\\
&-&(1 - z^2)(a\Omega)^2\bigg]\Theta = 0\;\;\;\;
\end{eqnarray}
to replace Eqs.~(\ref{eq:11}) and (\ref{eq:12}).  Here $C$ is the separation constant,
\begin{eqnarray}\label{eq:23}
y = r\!&+&\!r_+\frac{r_+ + r_-}{r_+ - r_-}\ln\Big|1 - \frac{r}{r_+}\Big|\nonumber\\
\!&+&\!r_-\frac{r_- + r_+}{r_- - r_+}\ln\Big|1 - \frac{r}{r_-}\Big|
\end{eqnarray}
tortoise coordinate in the Kerr metric \cite{Fiziev2}, applicable for arbitrary values of $r$, $y = 0$ when $r = 0$,
\begin{eqnarray}\label{eq:24}
r_\pm = \frac{r_s}{2}\pm\sqrt{\left(\frac{r_s}{2}\right)^2\!- a^2}
\end{eqnarray}
which are the two values (roots) of $r$ to let $\Delta = 0$ in Eq.~(\ref{eq:18}), marking the radii of the inner and outer horizons respectively.  Obviously, Eqs.~(\ref{eq:21}) and (\ref{eq:22}) reduce to Eq.~(\ref{eq:11}) and (\ref{eq:12}) when $a\rightarrow0$.  

\begin{figure}[t]
\centering\includegraphics[width=9cm]{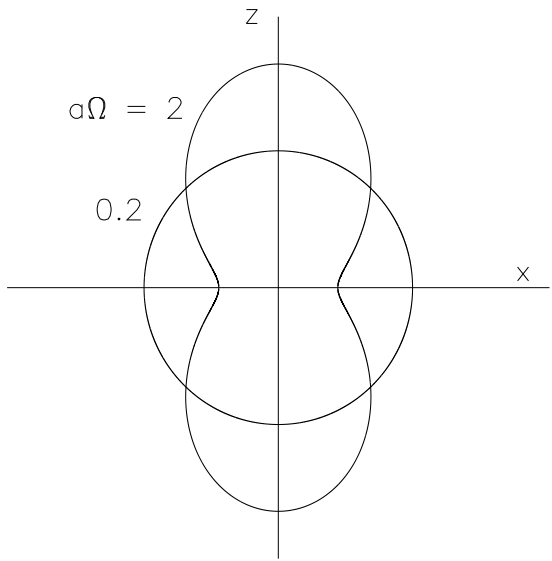}
\caption{Profile of the beam of radiation in the meridian plane found from Eq.~(\ref{eq:26}), areas of the profiles are normalized to be equal.}
\label{fig:3}
\end{figure}

Letting $C = 0$ in Eq.~(\ref{eq:21}) for radiation in the radial direction in the Kerr geometry, we find
\begin{eqnarray}\label{eq:25}
R(r) = \exp(-i\Omega y)
\end{eqnarray}
which resembles Eq.~(\ref{eq:14}) closely.  In FIG.~2 we plot the real part of $R(r)$ in Eq.~(\ref{eq:25}).  It is worth noting both the inner and outer horizons are within the Schwarzschild radius.  When the rate of rotation of the black hole reduces the inner horizon will move inwardly towards $r = 0$ but the outer horizon move outwardly towards $r = r_s$.  If we further let $m = 0$ in Eq.~(\ref{eq:22}) for radiation in the radial direction, then we have
\begin{eqnarray}\label{eq:26}
\Theta(\theta) = \cosh\big(a\Omega\cos\theta\big)
\end{eqnarray}
to modify the amplitude of the electromagnetic wave.  In FIG.~3 we plot normalized values of $\Theta(\theta)$ against $\theta$.  We have the explicit formula
\begin{eqnarray}\label{eq:27}
\eta = \Theta^2(0)/\Theta^2(\pi/2) = \cosh^2(a\Omega)
\end{eqnarray}
to measure the anisotropy of the beam, giving $\eta = 1.04$ ($a\Omega = 0.2$) and 14.15 ($a\Omega = 2$) for the two examples in FIG.~3.  Eq.~(\ref{eq:27}) tells us radiation in a Kerr black hole collimates along the axis of rotation, apparently because frame-dragging is weaker close to the axis and offers less impedance against beam propagation.  Fiziev pointed out radiation in the Kerr background could be in the form of bounded one-way running waves suitable to model astrophysical jets without a collimating magnetic field \cite{Fiziev2}.  Gariel and co-workers also discussed collimation by Kerr black holes, jet fueled by the Penrose process \cite{Gariel}.

\section{Other equations}\label{sec:4}
In the 1955 paper, Wheeler did not specify $e^\nu$ and $e^{-\lambda}$ in Eq.~(\ref{eq:9}) because he intended to let strong radiation modify the metric in Eq.~(\ref{eq:3}) to form self-sustaining entities called geons \cite{Wheeler}.  In 1957, Regge and Wheeler assumed weak perturbations to the Schwarzschild metric to investigate if $r_s$ is stable under the circumstance.  They found Eq.~(\ref{eq:11}), with an additional term $(6r_s/r^3)e^\nu$ in the brackets, known as the Regge-Wheeler equation \cite{Regge}.  The factor $r^2\sin^2\theta$ in Eqs.~(\ref{eq:5}) and (\ref{eq:6}) serves the purpose to remove the geometric effect of the ever-expanding front of the spherical wave, where its amplitude declines by the rule $1/r$ when $r\rightarrow\infty$. If we use
\begin{eqnarray}
xR(x)\;\;\;\;\mbox{and}\;\;\;\;\sqrt{1 - z^2\,}\,\Theta(z)\nonumber
\end{eqnarray}
to replace $R(x)$ and $\Theta(z)$ then we find from  Eqs.~(\ref{eq:11})
\begin{eqnarray}\label{eq:28}
\frac{1}{x^2}\frac{d}{dx}\left(x^2\frac{dR}{dx}\right) + \left(\Omega^2 - \frac{C}{r^2}e^\nu\!\right)\!R = 0
\end{eqnarray}
which reduces to the standard spherical Bessel equation in flat spacetime \cite{Abramowitz} as it should be after we have reinstalled the geometric effect of wavefront expansion.  We also find from Eq.~(\ref{eq:12})
\begin{eqnarray}\label{eq:29}
(1 - z^2)\frac{d^2\Theta}{dz^2} - 2z\frac{d\Theta}{dz} + \left[C + 1 - \frac{m^2 + 2}{1 - z^2}\right]\Theta = 0
\end{eqnarray}
which is the Legendre equation \cite{Abramowitz}.  It is worth noting the constant $C + 1$ in the above equation takes the value $\ell(\ell + 1)$, $\ell = 0, 1, 2, ...$, in the theory of Legendre functions so that in Eqs.~(\ref{eq:9}), (\ref{eq:10}), (\ref{eq:11}) and (\ref{eq:12}) we have $C = \ell^2 + \ell - 1$ instead of $C = \ell(\ell + 1)$ in \cite{Wheeler}. If we use
\begin{eqnarray}
yR(y)\;\;\;\;\mbox{and}\;\;\;\; \sqrt{1 - z^2\,}\,\Theta(z)\nonumber
\end{eqnarray}
to replace $R(y)$ and $\Theta(z)$ in Eqs.~(\ref{eq:21}) and (\ref{eq:22}) then we find two equations similar to Eqs.~(\ref{eq:28}) and (\ref{eq:29}), that reduce to the standard spherical Bessel equation and Legendre equation, respectively, in flat spacetime.  We can write the radial wave equation of Brill, Chrzanowski, Pereira, Fackerell and Ipser \cite{Brill} as:
\begin{eqnarray}\label{eq:30}
&&\frac{1}{r^2 + a^2}\,\frac{d}{dy}\left[\left(r^2 + a^2\right)\frac{dR}{dy}\right] +\nonumber\\
&&\;\;\;\;\;\;\;\;+ \left(\Omega^2 - \frac{C + a^2\Omega^2}{r^2 + a^2}\frac{\Delta}{r^2 + a^2}\right.\nonumber\\
&&\;\;\;\;\;\;\;\;+ \left.\frac{ma}{r^2 + a^2}\frac{ma - 2r_sr\Omega}{r^2 + a^2}\right)R = 0
\end{eqnarray}
which reduces to Eq.~(\ref{eq:28}) in flat spacetime. The second terms in Eqs.~(\ref{eq:21}) and (\ref{eq:30}) are almost identical apart from the extra constant, $a^2\Omega^2$, added to the separation constant $C$ in Eq.~(\ref{eq:30}).  We can write the angular wave equation in \cite{Brill} as:
\begin{eqnarray}\label{eq:31}
&&(1 - z^2)\frac{d^2\Theta}{dz^2} - 2z\frac{d\Theta}{dz}\nonumber\\
&&+\left[C - \frac{m^2}{1 - z^2} + m^2 + z^2(a\Omega)^2\right]\Theta = 0
\end{eqnarray}
which reduces to the Legendre equation when $a = 0$.

In 1889 Karl Heun introduced a second-order differential equation with four regular singular points which in the literature is usually presented in the following standard form \cite{Ronveaux}
\begin{eqnarray}\label{eq:32}
\frac{d^2\!H}{dw^2} + \left(\frac{\gamma}{w} + \frac{\delta}{w - 1} + \frac{\epsilon}{w - a}\right)\frac{dH}{dw}\nonumber\\
+ \frac{\alpha\beta w - q}{w(w - 1)(w - a)}H = 0
\end{eqnarray}
where $\gamma + \delta + \epsilon = \alpha + \beta + 1$.   It has exact solutions in the forms of local solutions (power series), Heun functions, Heun polynomials, and path-multiplicative solutions \cite{Ronveaux}.  In 1999 Suzuki, Takasugi and Umetsu found the Teukolsky equation, which is slightly more general than Eq.~(\ref{eq:30}), is a confluent limit of Eq.~(\ref{eq:32}) \cite{Suzuki}.  In 2006 Fiziev transformed both the Wheeler and Regge-Wheeler equations into Eq.~(\ref{eq:32}) in its confluent limit, with explicit presentations of $\alpha$, $\beta$, $\gamma$ and $\delta$ \cite{Fiziev}.

\section{velocity of light}\label{sec:5}
\begin{figure}[t]
\centering\includegraphics[width=9cm]{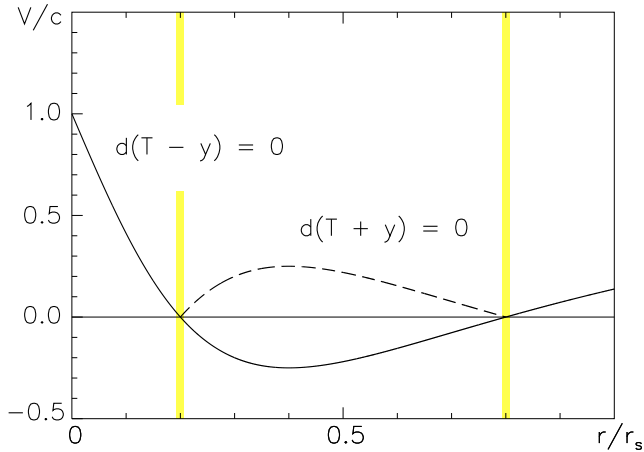}
\caption{Proper velocity of light in the radial direction of the Kerr background, with $\Omega r_s = 9.1$, $r_- = 0.2$ and $r_+ = 0.8$ in $r_s$.  The solid and dashed curves are from from $dy =\pm dT$ ($dy/dt =\pm c$) respectively.  The two gray strips mark locations of the inner and outer event horizons.}
\label{fig:4}
\end{figure}

In Eq.~(\ref{eq:14}) a constant value of $\exp(i\Omega T)R(r)$ marks the location of a wavefront.  It leads to a constant value of $T - x$ or $dx/dt = c$ so that, within the context of Eq.~(\ref{eq:14}), $dx$ amounts to a displacement measured locally and $dt$ the time elapsed on the local clock (proper time), albeit notationally $t$ suggests time from the clock with the distant observer.  We also have
\begin{eqnarray}\label{eq:33}
\frac{dr}{dt} = c\left(1 - \frac{r_s}{r}\right)
\end{eqnarray}
which must be the {\it proper velocity} of light in the radial direction of the Schwarzschild background because $dr$ is the displacement measured by the distant observer, but $dt$ still is from the local clock \cite{Ungar}.  In Eq.~(\ref{eq:33}) $dv/dr\rightarrow-\infty$, 0 and $c$ when $r\rightarrow0$, $r_s$ and $\infty$ respectively. Eq.~(\ref{eq:33}) tells us $dr/dt$ becomes negative when $r < r_s$ but that does not mean radiation suddenly reverses direction.  We can replace Eq.~(\ref{eq:14}) with $R(x) = \exp(i\Omega x)$  which too is a solution to Eq.~(\ref{eq:11}) when $C = 0$, leading through a constant value of $T + x$ to $dr/dt = c(r_s/r - 1)$. 

Similarly, we find  $dy/dt = c$ from Eq.~(\ref{eq:25}) with both $dy$ and $dt$ measured in the local frame taking into account the effects of both length contraction and time dilation.  We also have
\begin{eqnarray}\label{eq:34}
\frac{dr}{dt} = c\left(1 - \frac{r_s}{r + a^2/r}\right)
\end{eqnarray}
as the proper velocity of light in the radial direction in the Kerr background.  Note that in Eq.~(\ref{eq:34}) $dr/dt = 0$ is a quadratic equation with two roots ($= r_1$ and $r_2$).

In FIG.~4, we use the solid curve to present the values of the proper velocity of light, $dr/dt$ from Eq.~(\ref{eq:34}), in the radial direction of the Kerr background. It drops to become negative when $r_-\leq r\leq r_+$ and we use the dashed curve to present the values of $-dr/dt$ over the range.  We have $dr/dt = c$ when $r\rightarrow0$ and it remains the case as long as $a >0$.  However, when $a\rightarrow0$, the valley of the solid curve in FIG.~4 will become increasingly deeper and closer to $r = 0$ until Eqs.~(\ref{eq:33}) and (\ref{eq:34}), together with their solutions, become indistinguishable.

\section{space and time}\label{sec:6}
In a footnote in \cite{Fiziev}, Fiziev states that the variable $r$ plays the role of a time variable in the interior of Schwarzschild black holes.   According to his statement, the factor $\exp(i\Omega T)$ in Eq.~(\ref{eq:8}) marks the dependence of the solution on the new space variable.  We know from Eq.~(\ref{eq:33}) we have $v = dr/dt\rightarrow0$ when $r\rightarrow r_s$ as the proper velocity of light close to the event horizon from above it.  If $r$ and $t$ swap their roles in Eq.~(\ref{eq:33}), then we have $v = dt/dr\rightarrow\infty$ below the event horizon in sharp contrast to what happens above.  It is not clear how we should understand this result.

To see where the problem lies we recall $ds = 0$ for electromagnetic disturbances giving via Eqs.~(\ref{eq:1}) and (\ref{eq:3})
\begin{eqnarray}\label{eq:35}
\left(1 - \frac{r_s}{r}\right)c^2dt^2&=&\left(1 - \frac{r_s}{r}\right)^{-1}\!dr^2\nonumber\\
&+&r^2\Big(d\theta^2 + \sin^2\theta\,d\varphi^2\Big)
\end{eqnarray}
where the terms in $dt^2$ and $dr^2$ become negative when $r < r_s$.  To fix the problem, Fiziev swaps the two terms with a bold explanation that $r$ plays the role of time and $t$ the role of length, with the consequence of infinite velocity of light below the event horizon.

To avoid that consequence, we write Eq.~(\ref{eq:35}) as
\begin{eqnarray}\label{eq:36}
\left|1 - \frac{r_s}{r}\right|c^2dt^2&=&\left|1 - \frac{r_s}{r}\right|^{-1}\!dr^2\nonumber\\
&\pm&r^2\Big(d\theta^2 + \sin^2\theta\,d\varphi^2\Big)
\end{eqnarray}
where the sign in the second line is plus if $r > r_s$, otherwise it becomes minus.  Eq.~(\ref{eq:36}) also has a problem: its right-hand side may become negative when $r < r_s$, depending on the ratio between the first and second terms.  That ratio measures the angle of light rays against the radial direction.  The rays at the allowed angles form a cone, which is named the internal Oppenheimer-Snyder cone in \cite{Zheng}.  Since length contraction is the strongest ($g_{rr}$ largest) immediately below the event horizon, the angle of the cone is close to $\pi/2$ when $r\rightarrow r_s$ but 0 when $r\rightarrow0$.  We are reminded Eq.~(\ref{eq:36}) always applies to the solutions in Eqs.~(\ref{eq:14}) and (\ref{eq:25}) where the radiation is in the radial direction with $d\theta, d\varphi = 0$.

In 1958 Finkelstein proposed a different treatment for $r$ and $t$ in the Schwarzschild background \cite{Finkelstein}. He proposed to use
\begin{eqnarray}\label{eq:37}
T + r_s\ln\left(\frac{r}{r_s} - 1\right),\;\;\;\;\;\;\;\;r > r_s
\end{eqnarray}
to replace $T$ in Eq.~(\ref{eq:1}).  He announced the event horizon was no longer a singularity.  He claimed it would act as a perfect unidirectional membrane: causal influences can cross it but only in one direction.  Although Finkelstein did not state if Expression~(\ref{eq:37}) is applicable for $r < r_s$, he showed in FIG.~1 in \cite{Finkelstein} the interior of the black hole accommodates light cones, with both $r$ and $t$ playing the usual roles.  

In 1965 Penrose investigated gravitational collapse and spacetime singularities \cite{Penrose}. He applied Eq.~(\ref{eq:37}) and made it clear that a space-like sphere still is space-like after it collapses across the singularity at $r = r_s$ and becomes a trapped surface.  In FIG.~1 in the article, $r$ and $t$ always play their usual roles.  The angle of the future light cones diminishes the closer the distance to $r = 0$, consistent with our desire to let the right-hand side of Eq.~(\ref{eq:36}) be positive.

\section{running wave and standing wave}\label{sec:7}
In Eqs.~(\ref{eq:14}) and (\ref{eq:25}) $R(r)$ represents a harmonic wave running in the outward direction.  There must be a source consuming the mass of the black hole to generate radiation via some physics.  Otherwise, there must be a physical boundary to contain the radiation, similar to the metallic wall of a spherical microwave resonator.  The radiation will no longer be a single running wave (or running waves) but a standing wave made from two waves running in opposite directions.  For example, we have the following standing waves,
\begin{eqnarray}\label{eq:38}
\cos(\Omega x)\;\;\;\;\;\;\;\;\mbox{and}\;\;\;\;\;\;\;\;-\sin(\Omega x),
\end{eqnarray}
from the real and imaginary parts of Eq.~(\ref{eq:14}), both are summations of $\exp(\pm i\Omega x)$.  In Eq.~(\ref{eq:38}), the value of $\Omega$ must be discrete because a standing wave is from constructive interferences among its component running waves in multiple reflections.

But the event horizon is not a physical boundary.  In its place, there is nothing physical to reflect radiation.  Furthermore, an observer co-moving with the radiation forever sees the horizon ahead of him.   He will never know if the phase condition is correct for the radiation to interfere with itself constructively should reflection ever take place.  Indeed, when $r = r_s$, we find from Eq.~(\ref{eq:13}) $x = \infty$ which is not a definite coordinate for us to impose a boundary condition. 

In 2006, Fiziev made an attempt to present any disturbance to the Schwarzschild metric in terms of the quasi-normal modes (QNM) \cite{Fiziev}.  In this case, a QNM is a solution to the Regge-Wheeler equation subject to boundary conditions imposed at say $r = 0$ and $r_s$.  He did not establish a definite relation between the spectrum of the problem with the boundary conditions \cite{Fiziev}. 

In 2010 Fiziev made another attempt in the Kerr background.   He investigated Eq.~(\ref{eq:32}) in the form of the confluent limit of the Tuekolsky equation \cite{Fiziev2}.  He found numerous new exact solutions to the equation and clarified them into classes.  He paid specific attention to singular polynomial solutions and made solid progress.  He showed that a proper linear combination of such solutions, rather than QNM, could present bounded one-way running waves suitable as models of the observed astrophysical jets, consistent with our solution in Eqs.~(\ref{eq:25}), (\ref{eq:26}) and FIG.~3.  It is interesting that, in 2021, Horta\c{c}su also encountered difficulty when he attempted to calculate the reflective coefficients at the event horizons in the Kerr background \cite{Hortacsu}.

\section{Conclusions}\label{sec:8}
Fiziev stated the collimated one-way running waves from his composition indicate a universal mechanism in association with the pure gravitational field of rotating compact astrophysical objects of different nature \cite{Fiziev2}.  We have found an extension of the Wheeler equation in the Kerr background.  It enables us to specify the Fiziev running waves, together with an expression of the aspect ratio of the collimated beam in terms of the angular momentum of the celestial object and frequency of radiation.  We are a step closer to Fiziev's expectation for a more detailed mathematical development and a careful confrontation with the actual astrophysical observations against the above statement.

We have also clarified several conceptional issues.  We find the velocity of light in the tortoise coordinates is a proper velocity.  We prove space and time always play the usual roles across the Schwarzschild and Kerr metrics.  We show the event horizon is not a physical boundary suitable for a Fourier-style expansion.

\end{document}